\documentclass[twocolumn,showpacs,preprintnumbers,amssymb]{revtex4}
\usepackage{graphicx}
\usepackage{dcolumn}
\usepackage{amsmath}
\usepackage{amssymb}
\begin{document}
\preprint{APS/V2-05/10/04}

\title{The role of compressibility in solar wind plasma turbulence.}

\author{Bogdan Hnat}
 \email{hnat@astro.warwick.ac.uk}
\author{Sandra C. Chapman}
\author{George Rowlands}
 \affiliation{Physics Department, University of Warwick, Coventry, CV4 7AL, UK.}

\date{\today} % It is always today, date may be explicitly specified

\begin{abstract}
Incompressible Magnetohydrodynamics is often assumed to describe solar wind
turbulence. We use extended self similarity to reveal scaling in structure
functions of density fluctuations in the solar wind. Obtained scaling is then
compared with that found in the inertial range of quantities identified as
passive scalars in other turbulent systems. We find that these are not
coincident. This implies that either solar wind turbulence is compressible, or
that straightforward comparison of structure functions does not adequately
capture its inertial range properties.

\end{abstract}
\pacs{Valid PACS appear here}% PACS, the Physics and Astronomy
                             % Classification Scheme.
\keywords{scaling, passive scalar, solar wind, turbulence}
\maketitle
The supersonic and super-Alfv\'{e}nic flow of the solar wind offers a unique
opportunity to investigate the properties of the magnetized and turbulent
plasma. The transition from a laminar to turbulent flow requires large Reynolds
number $Re\!=\!LU/\nu$, and its magnetic counterpart $R_m\!=\!LU/\eta$, where $L$
is the energy injection scale length, $U$ is the velocity difference on scale
$L$, $\nu$ is the viscosity and $\eta$ is the magnetic diffusivity. Estimations
of the hydrodynamic and magnetic Reynolds numbers in the solar wind exceed
$10^{8}$\cite{verma,cho} compared with $Re \approx R_m \approx 10^4$ achieved in
direct numerical simulations (DNS)\cite{cattaneo,haugen,muller} and just few
hundreds in some magnetized liquid laboratory experiments \cite{paret,rutgers}.
The presence of turbulence in the solar wind is strongly suggested by numerous
observations. These include power law power spectra with $-5/3$ Kolmogorov-like
slopes in the kinetic and magnetic energy densities (eg., \cite{goldstein,cytu,matthaeus}) and non-Gaussian Probability Density Functions
(PDFs) (e.g., \cite{burlaga,marsch,valvo,hnat03}) found for fluctuations in
the velocity and the magnetic field.

These observations imply that solar wind turbulence shares many of its
statistical properties with incompressible isotropic hydrodynamic turbulence
\cite{kraichnan80,carbone,veltri}. As a result, the turbulent
dynamics of the solar wind is often modelled assuming that the plasma density
is constant. This assumption of incompressibility is particularly convenient
in analytical and numerical studies of magnetohydrodynamic (MHD) turbulence.
The assumption of incompressibility also appears to be in good agreement with
the results of compressive MHD simulations where generation of compressive modes
from Alfv\'{e}nic turbulence was found to be suppressed\cite{cho}.
\begin{figure}
\resizebox{0.675\hsize}{!}{\includegraphics{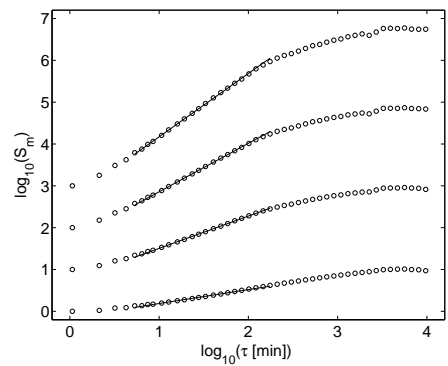}}
\caption{\label{fig1} Conditioned structure functions of fluctuations in the
density $\rho$ for slow solar wind.}
\end{figure}
In the context of the solar wind, incompressibility has been suggested to be a
reasonable approximation for plasma in fast wind streams\cite{bruno,tu90,valvo}. Considerable progress has been made by treating the solar wind as dominated by
Alfv\'{e}nic fluctuations (e.g.,\cite{kraichnan65,dobrowolny,goldreich,veli})
and with nearly incompressible magnetohydrodynamic theory\cite{zank}.

However, statistical features associated with turbulence have also been
identified in the density fluctuations derived from the solar wind observations.
A $k^{-5/3}$ scaling in the omnidirectional wave number spectrum of electron
and proton density has been reported as early as the $1970$s (see
\cite{montgomery} and the references therein). Models of the solar wind can
offer possible mechanisms for generating strong density fluctuations close to the
heliospheric current sheets\cite{malara}. Recent studies suggest that, at least
in the slow solar wind, the coupling of Alfv\'{e}nic and magnetosonic MHD modes
could not fully describe the nature of these fluctuations. For example,
the ratio $\delta \rho/\!\left<\rho \right>$ was found to be nearly constant and
independent from the amplitude of the magnetic field fluctuations $\delta B
/\!\left< B \right>$ in the slow solar wind\cite{spangler}.
More recently the density fluctuations were also found to exhibit non trivial,
approximately self-affine scaling\cite{hnat03} similar to that found in other
solar wind bulk plasma parameters. 

In this Letter we examine the scaling properties of the proton density $\rho$
in fast and in slow solar wind. We focus on a comparison of the scaling
properties of fluctuations in density with that of various passive scalars
identified in experiments, and direct numerical simulations, of fluid
turbulence. Previously, these passive scalars have been shown to exhibit scaling
that corresponds closely to that of the magnitude of the magnetic field $B$ in
the solar wind, although the data interval was not ordered by solar wind speed
\cite{bershadskii}. By assuming incompressibility it can be argued 
\cite{bershadskii} that $B$ should act as a passive scalar. We will for
completeness also examine the scaling properties of magnetic field magnitude in
fast and slow solar wind.

It is instructive to recall that the dynamics of a passive scalar
$T \equiv T(\mathbf{x},t)$ in a velocity field $\mathbf{v(\mathbf{x},t)}$ is
given by the advection equation
\begin{equation}
\partial_t T=-(\mathbf{v} \cdot \mathbf{\nabla}) T + \kappa \nabla^2  T.
\label{adv}
\end{equation}
where $\kappa$ is the diffusivity. Dynamical and statistical properties of a
passive scalar are more tractable analytically, as compared to active fields
like velocity, since the equation (\ref{adv}) is linear in $T$. In the case of
hydrodynamics, theoretical predictions have also been verified, to some extent,
experimentally using tracer particles that do not disturb the flow\cite{porta}.
\begin{figure}
\resizebox{0.675\hsize}{!}{\includegraphics{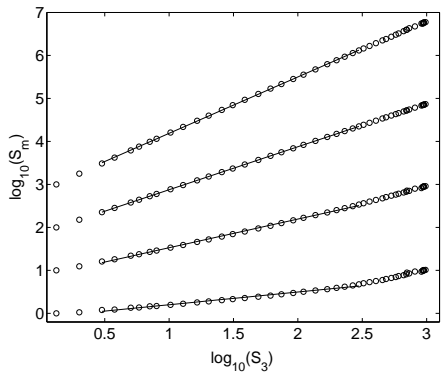}}
\caption{\label{fig2} ESS of conditioned structure functions of fluctuations in the density $\rho$ for slow solar wind.}
\end{figure}
\begin{figure}
\resizebox{0.675\hsize}{!}{\includegraphics{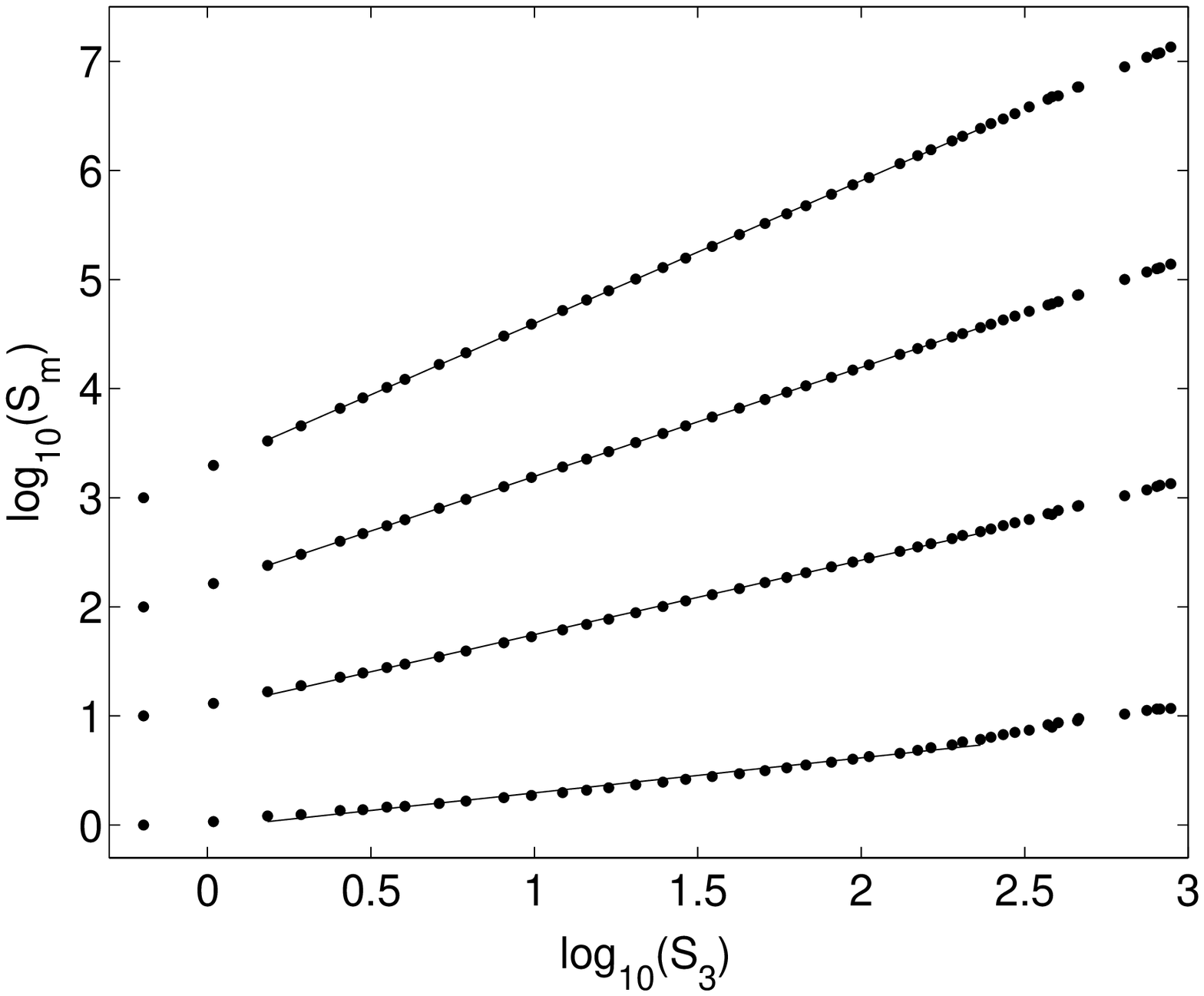}}
\caption{\label{fig3} ESS of conditioned structure functions of fluctuations in the density $\rho$ for fast solar wind.}
\end{figure}
Let us consider the compressible MHD equations for the evolution of the
magnetic field and the density:
\begin{equation}
\partial_t \mathbf{B} = \nabla \times (\mathbf{v} \times \mathbf{B})+\eta \nabla^2 \mathbf{B} \quad \textrm{and} \quad \partial_t \rho + \mathbf{\nabla} \cdot (\mathbf{v} \rho) = 0.
\label{mhdcomp}
\end{equation}
Given the assumption of incompressibility ($\nabla \cdot \mathbf{v} = 0$), it
immediately follows that $\rho$ should behave as a passive scalar in the
turbulent solar wind flow:
\begin{equation}
\partial_t \rho = -(\mathbf{v} \cdot \mathbf{\nabla}) \rho.
\label{mhdRincomp}
\end{equation}

To investigate the scaling properties of the density $\rho$ and the magnetic
field magnitude $B$ we use  $64$ seconds averaged data from the ACE spacecraft \cite{ace} set spanning from $01/01/1998$ to $12/31/2001$. This interval 
includes dates previously considered in Ref.\cite{bershadskii}. The slow and fast
solar wind are known to exhibit distinct phenomenology (e.g., 
\cite{bruno,valvo}).
We thus split the data into slow and fast solar wind sets using $450$km/s wind
speed as a separation criteria. The resulting data sets consist of $\sim 1
\times 10^6$ samples for the slow wind and $\sim 0.6 \times 10^6$ for the fast
wind. We apply structure function\cite{stolovitzky} analysis to fluctuations in
density for slow and fast wind separately. Generalized structure functions
$S_m$ of fluctuations in say, the density $\rho(t)$, on timescale $\tau$ are
defined as moments $m$ through $S_m(\tau)\!=\! \left<|\rho(t+\tau)-\rho(t)|^m \right>$ where the ensemble $\left< \ldots \right>$ is taken in the time domain
\cite{frisch}. If scaling is present in the time series we expect these to show
a power law dependence on the temporal scale $\tau$, i.e.,
$S_m \propto \tau^{\zeta(m)}$.

Finite, experimental data sets  include a small number of extreme events
(outliers) that, due to poor statistics, may obscure the correct scaling
of the high order moments. Here, we will exclude these events by the use of
{\it conditioning} \cite{veltri}. This approach puts a limit on the range of
fluctuations used in computing structure functions. This limit is varied with
the temporal scale $\tau$ to account for the growth of range with temporal scale
in the signal. In our case we defined this threshold as $15 \sigma(\tau)$, where
$\sigma(\tau)$ is a standard deviation of fluctuations on temporal scale $\tau$.
\begin{figure}
\resizebox{0.675\hsize}{!}{\includegraphics{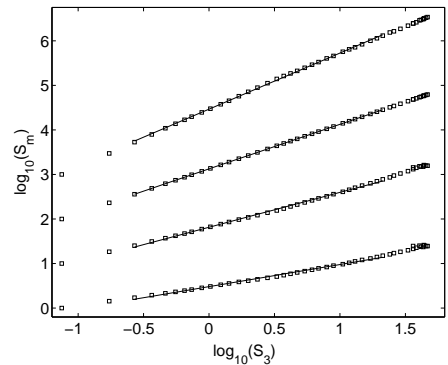}}
\caption{\label{fig4} ESS of conditioned structure functions of fluctuations in the magnetic field magnitude $B$ for slow solar wind.}
\end{figure}
\begin{figure}
\resizebox{0.675\hsize}{!}{\includegraphics{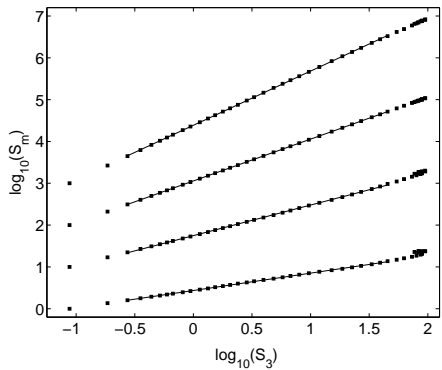}}
\caption{\label{fig5} ESS of conditioned structure functions of fluctuations in the magnetic field magnitude $B$ for fast solar wind.}
\end{figure}
We stress that conditioning improves the scaling where it already exists but
does not enforce it on the investigated data, if the applied threshold is
sufficiently large. In practice, for the limit chosen here, we eliminate less
then $1\%$ of the data points.

Figure \ref{fig1} shows the structure functions $S_m$ plotted versus $\tau$ on
logarithmic axes for orders $1 \leq m \leq 4$ for fluctuations in density in
the slow solar wind. The plot shows a scaling region extending from
$\tau \sim10$ minutes to $\tau \sim3$ hours ($\sim1.5$ decades on the
logarithmic scale). The quality of the scaling deteriorates when fast wind
streams are considered. The scaling regions can be still identified but they
extend only $<1$ decade from $\sim10$ minutes to $1$ hour.

It has been empirically shown that Extended Self Similarity (ESS) approach can
considerably extend the region of scaling in structure functions\cite{benzi}.
The method is based on the assumption that $S_m(\tau)\propto S_p^{\eta(m)}(\tau)$. This suggests that scaling should emerge when the quantity
$S_m$ is plotted as a function of $S_p$. The scaling exponent $\zeta$ can then
be obtained straightforwardly from the relation $\zeta(m)=\zeta(p) \eta(m)$.
For the density fluctuations in the solar wind, we find that $S_3$ is close to
unity and we plot $S_m$ versus $S_3$ on logarithmic axes for fluctuations in the
density in slow and fast solar wind, in figures \ref{fig2} and \ref{fig3}
respectively. These figures demonstrate that we can extend scaling in the
density in both the slow and the fast wind streams to over $2$ decades when ESS
is applied.

ESS was previously applied to the magnitude of magnetic field in an
undifferentiated interval of solar wind\cite{bershadskii} and for completeness
we give the analysis in slow and fast wind here. Figures \ref{fig4} and
\ref{fig5} show $S_m$ versus $S_3$ on logarithmic axes for fluctuations in
the magnitude of magnetic field. Previously, the scaling exponents were
obtained by considering $S_m$ versus $S_4$\cite{bershadskii} and we have 
verified that these closely correspond to the values found here.
The local slopes of the third order structure functions $S_3$ then yield an
estimate of the exponent $\zeta(3)$ and these are detailed in Table \ref{tab1}.

\begin{figure}
\resizebox{0.675\hsize}{!}{\includegraphics{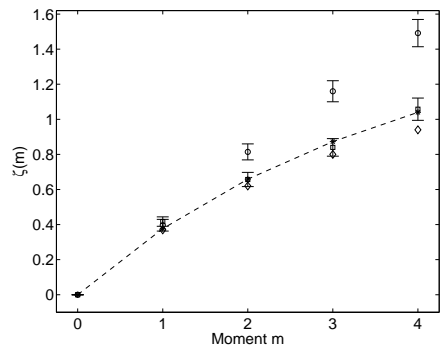}}
\caption{\label{fig6} Scaling exponents of the conditioned
structure functions for $\delta B$ (squares) and $\delta \rho$
(circles) in slow solar wind. Dashed line and diamonds correspond
to DNS\cite{watanabe} and experimental\cite{chavarria} results,
respectively.}
\end{figure}

\begin{figure}
\resizebox{0.675\hsize}{!}{\includegraphics{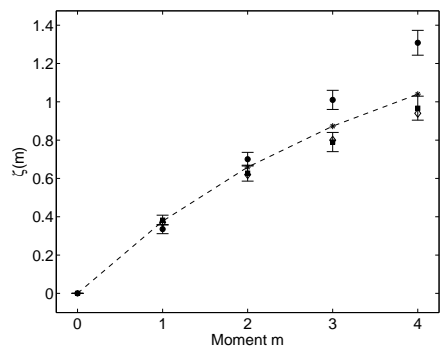}}
\caption{\label{fig7} Scaling exponents of the conditioned structure functions
for $\delta B$ (squares) and $\delta \rho$ (circles) in fast solar wind. Other
symbols same as in figure \ref{fig6}.}
\end{figure}

We can now directly compare the scaling found for the density $\rho$ with that
identified for quantities acting as passive scalars and that of the magnetic
field magnitude $B$. The resulting functional form of the scaling exponents
$\zeta(m)$ for the density and magnetic field magnitude in slow and fast solar
wind are shown in figures \ref{fig6} and \ref{fig7} respectively. For 
comparison, the scaling exponents obtained for passive scalars from the DNS 
\cite{watanabe} and the the wind tunnel experiment\cite{chavarria} are also
shown.
We immediately see that whereas the exponents for magnetic field magnitude and
the passive scalars fall close to each other on these plots (as also reported
in \cite{bershadskii}) they are distinct from those obtained for the solar wind
density in both slow and fast solar wind. The fluctuations in $\delta B$, for
slow and fast wind, and that of the passive scalars exhibit multi-fractal
scaling. Intriguingly, the density fluctuations are nearly self-affine with
scaling exponent $\alpha$ that differs in slow and fast wind. The values of
these exponents are: $\alpha^{\rho}_{slow}=0.39\pm0.03$ and
$\alpha^{\rho}_{fast}=0.33\pm0.03$.

This suggests one of two possible conclusions. The first of these is that the
turbulent solar wind is compressible and equation (\ref{mhdRincomp}) does not
hold, so that the density is an active quantity. This is rather surprising and
calls into question the significant body of work on MHD turbulence in the solar
wind that relies on the assumption of incompressibility (for example,
\cite{kraichnan65,dobrowolny,goldreich,veli}). This also invalidates the arguments in Ref.\cite{bershadskii} required to cast the advection equation for 
$B$ in the form of a passive scalar which require incompressibility, yielding:
\begin{equation}
\partial_t B=-(\mathbf{v} \cdot \mathbf{\nabla})B + \eta \nabla^2 B + \lambda B,
\label{mhdBincomp}
\end{equation}

The second possibility is that commonality of scaling (as measured by structure
functions) does not imply shared phenomenology, and as a corollary, an absence
of such a commonality does not imply distinct phenomenology.
The fact that several quantities share the same structure functions is then
coincidental rather than expressing some universality of fluid turbulence.
This implies that the structure functions of a single quantity do not fully
capture the phenomenology of a given turbulent system. This has implications
for analysis of turbulence which has been achieved through comparison between
(multi-fractal) models of turbulence and the data, via the structure functions
(see, for example, \cite{frisch}).
\begin{table}[t]
\begin{center}
\caption{Exponents $\zeta(3)$ obtained from figures \ref{fig3}.}
\begin{tabular}{cccc}
\hline
\hline
Quantity        &Slow/Fast    & $\zeta(3)$ & Range [min]\\
\hline
$\delta |B|$&Slow&$0.85\pm0.06$& $10-500$\\
$\delta \rho$&Slow&$1.16\pm0.08$& $10-120$\\
$\delta |B|$&Fast&$0.79\pm0.05$& $8-200$\\
$\delta \rho$&Fast&$1.01\pm0.07$& $6-100$\\
\hline
\end{tabular}
\label{tab1}
\end{center}
\end{table}

A possible resolution  may be found in the form of equations (\ref{mhdRincomp})
and (\ref{mhdBincomp}). It is known that the molecular diffusivity can change
the inertial range scaling properties of a passive scalar\cite{kraichnan94}.
The observed differences could then be due to the presence of the diffusive term
in the magnetic field equation (\ref{mhdBincomp}) and its absence in the
density equation (\ref{mhdRincomp}). This may account for the different scaling found in these quantities.

Intriguingly, equation (\ref{mhdRincomp}), when written for the moments of
fluctuations in $\rho$, has no explicit dependence on the order of the moment
\cite{kraichnan94}
\begin{equation}
[\partial_t + \mathbf{v(\mathbf{x},t)} \cdot \mathbf{\nabla} + \mathbf{v(\mathbf{x'},t)} \cdot \mathbf{\nabla'}] (\delta \rho)^n = 0,
\label{mhdstrf}
\end{equation}
where $(\delta \rho)^n \equiv [\rho(\mathbf{x},t)-\rho(\mathbf{x'},t)]^n$.
In this case fluctuations in density should be simply those imposed on the
initial condition $\delta \rho(\mathbf{x},0)$. These may be mediated via large
scale coherent structures (shocks and coronal mass ejections). This may suggest
a solar origin of these fluctuations in the density. It may thus be informative
to attempt to relate the scaling found in the density fluctuations in the
solar wind, with that of the solar corona. The power law scaling of X-ray flux
from solar flares\cite{aschwanden} is intriguing in this regard.

\section{Acknowledgment}
S. C. Chapman and B. Hnat acknowledge support from the PPARC and G. Rowlands
from the Leverhulme Trust. We thank the ACE SWEPAM instrument team and the ACE
Science Center for providing the ACE data.
%%%%%%%%%%%%%%%%%%%%%%%%%%%%%%%%%%%%%%%%%%%%%%%%%%%%%%%%%%%%%%%%%%%%%%


\begin{thebibliography}{37}
\bibitem[1]{verma} M.~K.~Verma, J. Geophys. Res. {\bf 101}, 27543-27548 (1996).
\bibitem[2]{cho} J~.Cho and A.~Lazarian, Phys. Rev. Lett. {\bf 88}, 245001
(2002).
\bibitem[3]{cattaneo} F.~Cattaneo, T.~Emonet, and N.~Weiss, ApJ. {\bf 588},
1183--1198 (2003).
\bibitem[4]{haugen} N.~E.~L.~Haugen and A.~Brandenburg, Phys. Rev. E {\bf 70},
026405 (2004).
\bibitem[5]{muller} W.-C.~M\"{u}ller, D.~Biskamp, Phys.~Rev.~Lett. {\bf 84}(3), 475 (2000).
\bibitem[6]{paret} J.~Paret and P.~Tabeling, Phys. Rev. Lett. {\bf 79}, 4162
(1997).
\bibitem[7]{rutgers} M.~A.~Rutgers, Phys. Rev. Lett. {\bf 81}, 2244-2247 (1998).
\bibitem[8]{goldstein}  M.~L.~Goldstein and D.~A.~Roberts, Phys. Plasmas
{\bf 6}, 4154--4160 (1999).
\bibitem[9]{cytu} C.-Y.~Tu and E.~Marsch, Space Sci. Rev. {\bf 73}, 1--210 (1995).
\bibitem[10]{matthaeus}  W~.H.~Matthaeus and M.~L.~Goldstein, J. Geophys. Res {\bf 87}, 6011 (1982).
\bibitem[11]{burlaga} L.~F.~Burlaga, J. Geophys. Res. {\bf 106}, 15,917--15,927 (2001).
\bibitem[12]{marsch} E.~Marsch and C.-Y.~Tu, Ann. Geophysicae {\bf 12}, 1127 (1994).
\bibitem[13]{valvo} L.~Sorriso-Valvo, V.~Carbone, P.~Giuliani, P.~Veltri, 
R.~Bruno, V.~Antoni and E.~Martines, Planet. Space Sci. {\bf 49}, 1193--1200
(2001).
\bibitem[14]{hnat03} B.~Hnat, S.~C.~Chapman, G.~Rowlands, Phys.~Rev.~E
{\bf 67}, 056404 (2003).
\bibitem[15]{kraichnan80}  R~.H.~Kraichnan and D.~Montgomery, Rep. Prog. Phys.
{\bf 43}, 547 (1980).
\bibitem[16]{carbone} V.~Carbone, P.~Veltri, and R.~Bruno, Phys. Rev. Lett.
{\bf75}, 3110--3113 (1995).
\bibitem[17]{veltri} P.~Veltri, Plasma Phys. Control. Fusion {\bf 41},
A787--A795 (1999).
\bibitem[18]{bruno} R.~Bruno, V.~Carbone, L.~Sorriso-Valvo, B.~Bavassano,
J. Geophys. Res. {\bf 108}, 1130, 10.1029/2002JA009615 (2003).
\bibitem[19]{tu90} C.-Y.~Tu, E.~Marsch, H.~Rosenbauer, Geophys. Res. Lett. {\bf 17}, 283 (1990).
\bibitem[20]{kraichnan65}  R~.H.~Kraichnan, Phys. Fluids {\bf 8}, 1385-7 (1965).
\bibitem[21]{dobrowolny} M.~Dobrowolny, A.~Mangeney, and P.~L.~Veltri,
Phys. Rev. Lett. {\bf 45}, 144--147 (1980).
\bibitem[22]{goldreich} P.~Goldreich and H.~Sridhar, Astrophys. J. {\bf 438}, 763 (1995).
\bibitem[23]{veli} M.~Velli, R.~Grappin, and A.~Mangeney, Phys. Rev. Lett.
{\bf 63}, 1807--1810 (1989).
\bibitem[24]{montgomery}  D.~Montgomery, M.~R.~Brown, W~.H.~Matthaeus, J. 
Geophys. Res {\bf 92}, 282--284 (1987).
\bibitem[25]{spangler} S.~R.~Spangler and L.~G.~Spitler, Phys. of Plasmas
{\bf 11}, 1969-1977 (2004).
\bibitem[26]{zank} G.~P.~Zank, W~.H.~Matthaeus, Phys. Rev. Lett. {\bf 64}, 1243 (1990).
\bibitem[27]{malara} F.~Malara, P.~Veltri, L.~Primavera, Phys. Rev. E {\bf 56}, 
3508-3514 (1997).
\bibitem[28]{bershadskii} A.~Bershadskii and K.~R.~Sreenivasan, Phys. Rev. Lett.
{\bf 93}, 064501 (2004).
\bibitem[29]{porta} A.~La Porta, G.~A.~Voth, A.~M.~Crawford, J.~Alexander, E.~Bodenschatz, Nature {\bf 409}, 1017--1019 (2001). 
\bibitem[30]{ace} E.~C.~Stone {\it et~al.}, Space Sci. Rev. {\bf 86}, 1 (1998). 
\bibitem[31]{stolovitzky} G.~Stolovitzky, K.~R.~Sreenivasan, and A.~Juneja,
Phys. Rev. E {\bf 48}, R3217-R3220 (1993).
\bibitem[32]{frisch} U.~Frisch, {\em Turbulence. The legacy of  A.N. Kolmogorov}
(Cambridge University Press, Cambridge, 1995).
\bibitem[33]{benzi} R.~Benzi, S.~Ciliberto, R.~Tripiccione, C.~Baudet,
F.~Massaioli, and S.~Succi, Phys. Rev. E {\bf 48}, R29-R32 (1993).
\bibitem[34]{watanabe} T.~Watanabe and T.~Gotoh, New Journal of Physics {\bf 6}, 40 (2004).
\bibitem[35]{chavarria} G.~Ruiz-Chavarria, C.~Baudet, S.~Ciliberto, Physica D
{\bf 99}, 369-380 (1996).
\bibitem[36]{kraichnan94} R~.H.~Kraichnan, Phys. Rev. Lett. {\bf72}, 1016-1019 (1994).
\bibitem[37]{aschwanden} M.~J.~Aschwanden and C.~E.~ Parnell, ApJ. {\bf 572},
1048-1071 (2002).
\end{thebibliography}
\end{document}